\documentclass[useAMS,usenatbib]{mn2e}

\usepackage{graphicx}
\usepackage{times}
\usepackage{amssymb}
\usepackage{amsmath}

\title[Calculating gravitational waves from white dwarf binaries]{On
  the point mass approximation to calculate the gravitational wave
  signal from white dwarf binaries} \author[van den Broek, Nelemans, Dan \& Rosswog]{D. van den Broek$^{1}$\thanks{E-mail:
  DickvandenBroek@student.ru.nl; nelemans@astro.ru.nl},
  G. Nelemans$^{1,2,3}$, M. Dan$^{4}$ and S. Rosswog$^{4}$\\ 
$^{1}$Department of Astrophysics/IMAPP, Radboud  University Nijmegen, P.O. Box 9010, NL-6500 GL, The Netherlands\\
$^{2}$Institute for Astronomy, KU Leuven, Celestijnenlaan 200D, 3001 Leuven, 
Belgium\\
$^{3}$Nikhef, Science Park 105, 1098 XG Amsterdam, The Netherlands\\
$^{4}$School of Engineering and Science, Jacobs University Bremen, Campus Ring 1, 28759 Bremen, Germany}
\begin{document}

\date{\today}

\pagerange{\pageref{firstpage}--\pageref{lastpage}} \pubyear{2012}

\maketitle

\label{firstpage}

\begin{abstract}
Double white dwarf binaries in the Galaxy dominate the gravitational
wave sky and would be detectable for an instrument such as LISA.  Most
studies have calculated the expected gravitational wave signal under
the assumption that the binary white dwarf system can be represented
by two point masses in orbit. We discuss the accuracy of this
approximation for real astrophysical systems. For non-relativistic
binaries in circular orbit the gravitational wave signal can easily be
calculated. We show that for these systems the point mass
approximation is completely justified when the individual stars are
axisymmetric irrespective of their size. We find that the signal
obtained from Smoothed-Particle Hydrodynamics simulations of tidally
deformed, Roche-lobe filling white dwarfs, including one case when an
accretion disc is present, is consistent with the point mass
approximation. The difference is typically at the level of one per
cent or less in realistic cases, yielding small errors in the inferred
parameters of the binaries.
\end{abstract}


\section{Introduction}
At low frequencies (mHz), millions of double white dwarf binaries in
the Galaxy are expected to dominate the gravitational wave (GW)
sky. At the lowest frequencies they form an unresolved foreground,
while at frequencies above several mHz, thousands of sources would be
individually detectable for an instrument such as the Laser
Interferometer Space Antenna, LISA (\citealt{e1}, \citealt{l1},
\citealt{h1}, \citealt{nyp01, nyp03}) or eLISA/NGO
(\citealt{eLISA}). These binaries come in two flavours: detached
systems and semi-detached (mass-transferring) systems that are know as
AM CVn systems (see \citealt{solheim2010}, \citealt{marsh2011} for
reviews). Several known binaries should be detected by LISA within the
first weeks of operation and are known as verification binaries
(\citealt{sv2006}, \citealt{rgb+07, roelofs2010},
\citealt{brown2011}). By measuring their gravitational wave amplitude
and frequency (evolution), the type of astrophysical source and its
parameters can be determined (e.g. \citealt{cutler1994},
\citealt{littenberg2011}, \citealt{blaut2011}).  In all these
calculations the gravitational wave signal was determined under the
assumption that the binary white dwarf system can be represented by
two point masses in orbit, even for the tidally deformed stars in
semi-detached binaries. The goal of this study is to determine the
accuracy of this assumption. In section \ref{section gravwaves} we
will discuss our method of calculating the GW signal. In section
\ref{section baloons} we will give an algebraic view on the assumption
of using point masses and in section \ref{section SPH} we will
calculate the GW signal from smoothed-particle hydrodynamics (SPH)
simulations of AM CVn stars. In section \ref{section conclusions} we
discuss the conclusions of this study.

\section{Gravitational waves from arbitrary sources in circular motion}\label{section gravwaves}

We calculate the GW wave signal from a collection of point particles
with arbitrary coordinates and masses, all rotating about a fixed
point with the same angular speed $\omega$. Since the stars do not
move at highly relativistic speeds we use linearised general
relativity. In linearised general relativity the trace reversed metric
for any non-relativistic, far away source is given by
\citep[e.g.][]{rindler}
\begin{equation}
 \overline{h}_{ij} = \frac{-2G}{c^4R}\int \frac{d^2}{dt^2}\rho \bmath{x_ix_j}dV = \frac{-2G}{c^4R}\sum_\alpha \frac{d^2}{dt^2}m_\alpha \bmath{x_{i\alpha} x_{j\alpha}},
\end{equation}
where R is the distance from observer to the source taken as an
average over the distance to all source points. $m_\alpha$ and
$\bmath{x_{i\alpha}}$ are the mass and coordinates of points in the
source. $\sum_\alpha m_\alpha \bmath{x_{i\alpha} x_{j\alpha}}$ is the
quadrupole moment of the source. In the transverse-traceless (TT)
gauge for a wave travelling in the z-direction, the wave has only two
degrees of freedom left. They manifest themselves as so called $+$ and
$\times$ polarisations that can be measured by a detector. The wave
metric then becomes \citep{rindler}
\begin{equation}
h_{ij}=\
  \left(
  \begin{array}{ccc}
   h_{+} & h_{\times} & 0\\
   h_{\times} & -h_{+} & 0\\
   0 & 0 & 0
  \end{array}
  \right).
\end{equation}
$h_+$ and $h_\times$ can be obtained by \citep{p1}
\begin{equation}
h_+ = \frac{1}{2}(\overline{h}_{xx}-\overline{h}_{yy})\\
h_\times = \overline{h}_{xy}.
\end{equation}
If we now take $N$ particles and assign them masses $m_\alpha$ and positions in
polar coordinated $r_\alpha$ and $\phi_\alpha = \omega
t+\theta_\alpha$, we can derive a general expression for $h_+$ and
$h_\times$. Also taking into account the inclination angle $i$ for the
relative orientation of the source, we get:
\begin{equation}\label{h+}
 h_+  =      \frac{-4\omega^2G}{c^4 R}(1+\cos^2 i) \left(S_1\cos2\omega t -S_2\sin2\omega t \right)
\end{equation}
\begin{equation}\label{hx}
 h_\times =  \frac{-4\omega^2G}{c^4 R} \cos i      \left(S_2\cos2\omega t -S_1\sin2\omega t \right),
\end{equation}
where
\begin{equation}\label{S1}
S_1 \equiv \sum_\alpha m_\alpha r^2_\alpha \cos2\theta_\alpha
\end{equation}
\begin{equation}\label{S2}
S_2 \equiv \sum_\alpha m_\alpha r^2_\alpha \sin2\theta_\alpha.
\end{equation}
These expressions depend only on the given initial coordinates and can
easily be calculated numerically. Because $h_+$ and $h_\times$ have a
cosine and a sine term with different amplitudes it is useful to
define the average strain amplitude as:
\begin{equation}\label{strainamp}
 h \equiv \sqrt{\frac{1}{2}(h_{+max}^2+h_{\times max}^2)}.
\end{equation}
As we will later determine the accuracy of the point mass
approximation, it is worth estimating the error we make by using the
linearised theory for these systems. The first Post-Newtonian
correction is proportional to $(v/c)^2$
(e.g. \citealt{1973grav.book.....M},\citealt{blanchet1995}) which for
double white dwarf systems with orbital period of several minutes is
of order $10^{-5}$.

\section{Parallel axis theorem for gravitational waves}\label{section baloons}

Using equations (\ref{h+}) and (\ref{hx}) we look at what we can say
about the two point mass approximation algebraically. The parts of the
equations that depend on the configuration of the system are $S_1$ and
$S_2$. For these expressions something very similar to the parallel
axis theorem for moment of inertia can be expressed. Consider a body
in the $xy$-plane with its centre of mass at the origin.  Its mass
distribution can be described by point masses with coordinates
$x_\alpha$,$y_\alpha$ and mass $m_\alpha$ with respect to its own
centre of mass. So $S_1$ and $S_2$ are obtained from
Eqs.~\ref{S1},\ref{S2}. We call these $S_{\rm 1, spinning}$ and
$S_{\rm 2, spinning}$, because these are the $S$'s that would arise if
the body was spinning around its centre of mass. We now change the
coordinates of our origin to another point in the plane such that all
point masses get new coordinates $x'_\alpha = x_\alpha - x_{\rm CM}$ and
$y'_\alpha = y_\alpha - y_{\rm CM}$. All coordinates denoted with
subscript CM refer to the location of the body's centre of mass in the
new frame and we assume the body is in circular motion around the new
origin. 

To do the translation we first transform to Cartesian coordinates,
translate and then transform back to polar coordinates. For the
transform we use 
\begin{equation}
\frac{y_\alpha}{x_\alpha} = \left\{ \begin{array}{l l}
                                     \tan(\theta_\alpha) &    \text{for} \quad x_\alpha\geq0\\
                                     \tan(\theta_\alpha+\pi) &\text{for} \quad x_\alpha<0\\
                                    \end{array} \right \}.\\
\end{equation}
 The mirror symmetry, $\cos(2\theta) = \cos(2(\theta+\pi))$ and
 $\sin(2\theta) = \sin(2(\theta+\pi))$ , relieves us from having to
 split the sum over particles into separate sums for positive and
 negative $x_\alpha$, allowing the new $S$ to be written as the old
 $S$ plus an additional term. The resulting $S$'s that determine the GW
 radiation are: 
\begin{equation}\label{S_1 translated}
S_1 = S_{\rm 1, spinning}+\sum_\alpha m_\alpha  r^2_{\rm CM}\cos2\theta_{\rm CM}
\end{equation}
\begin{equation}\label{S_2 translated}
S_2 = S_{\rm 2, spinning}+\sum_\alpha m_\alpha  r^2_{\rm CM}\sin2\theta_{\rm CM},
\end{equation}
i.e. a first term denoting the original $S_{\rm 1, spinning}$ and
$S_{\rm 2, spinning}$ and a
second displacement term that sums all mass in one point, so the
change of coordinates has contributed only the effect of a point mass
at the body's centre of mass.  

So far we have considered the case in which there is one body rotating
around an arbitrary point. In the case of a binary system, the
procedure can be repeated for the second star and the $x_{\rm CM}, y_{\rm CM}$
for each star now refer to the distance of the centres of the two
stars to the system barycentre.  

We thus can conclude that, when determining $h$, every body with
$S_{\rm 1, spinning} = S_{\rm 2, spinning} = 0$ can be considered as a
point mass without implications to the result. These are bodies that
do not radiate GW if they were only spinning around an axis through
their own CM. In other words: bodies without a quadrupole moment, like
axisymmetric spheres or disks. This result even holds when the body
overlaps with the point it is rotating about, if that is physically
possible. So only asymmetries in the stars that form a binary GW
source may lead to deviations from the result as calculated using a
source consisting only of two point masses.

\section{Numerical calculation of gravitational waves from SPH simulations}\label{section SPH}

For semi-detached binaries, we know they do not consist of spherical
stars. The contribution of the accretion disk will depend on if it is
circular or not and on how the mass is distributed over the disk. The
contribution of the donor will also depend on the mass distribution
over its shape. We therefore need to look at the GW signals of more
realistic mass distributions.

\subsection{The simulations}

\begin{figure}
\includegraphics[keepaspectratio, height = \linewidth,angle=-90,clip=]{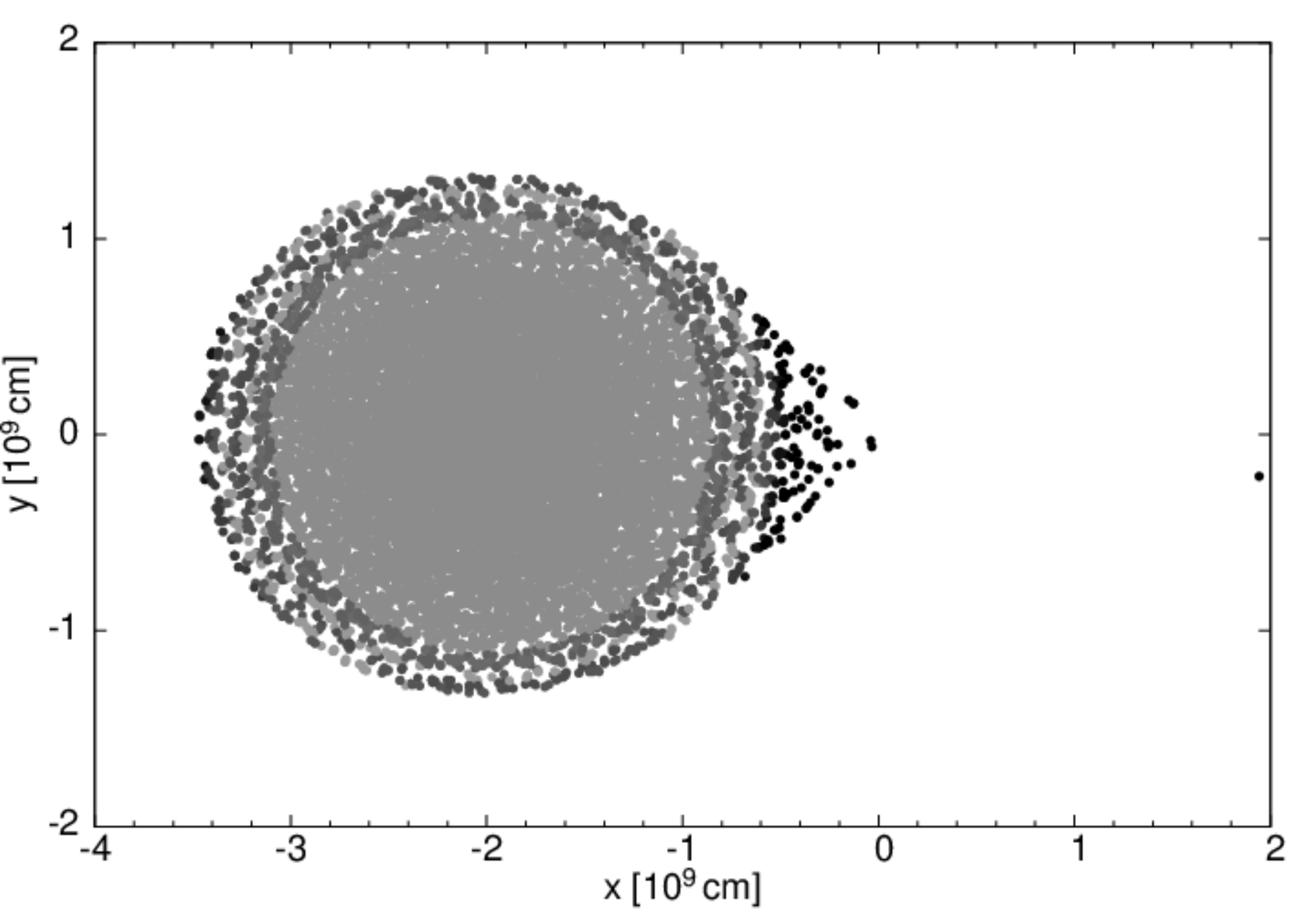}
\caption{\label{rocheplot} Projection of SPH particles onto the
  orbital plane of one of the stars in a double white dwarf system
  with $M_{\rm donor}=M_{\rm accretor} = 0.2 M_{\odot}, q =1$. The
  colour coding accentuates the asymmetry in the mass distribution
  that is responsible for the deviation of the gravitational wave
  signal from that of two point masses. For each ring around the
  center of mass of the star the ratio of the mass on the left ($x$
  coordinates smaller than $x_{\rm CM}$) to that on the right($x$
  coordinates larger than $x_{\rm CM}$) is calculated. Light grey
  represents values of unity (smoothed for clarity) and black values below
  0.5.}
\end{figure}

\begin{figure}
\includegraphics[keepaspectratio, height = \linewidth,angle=-90,clip=]{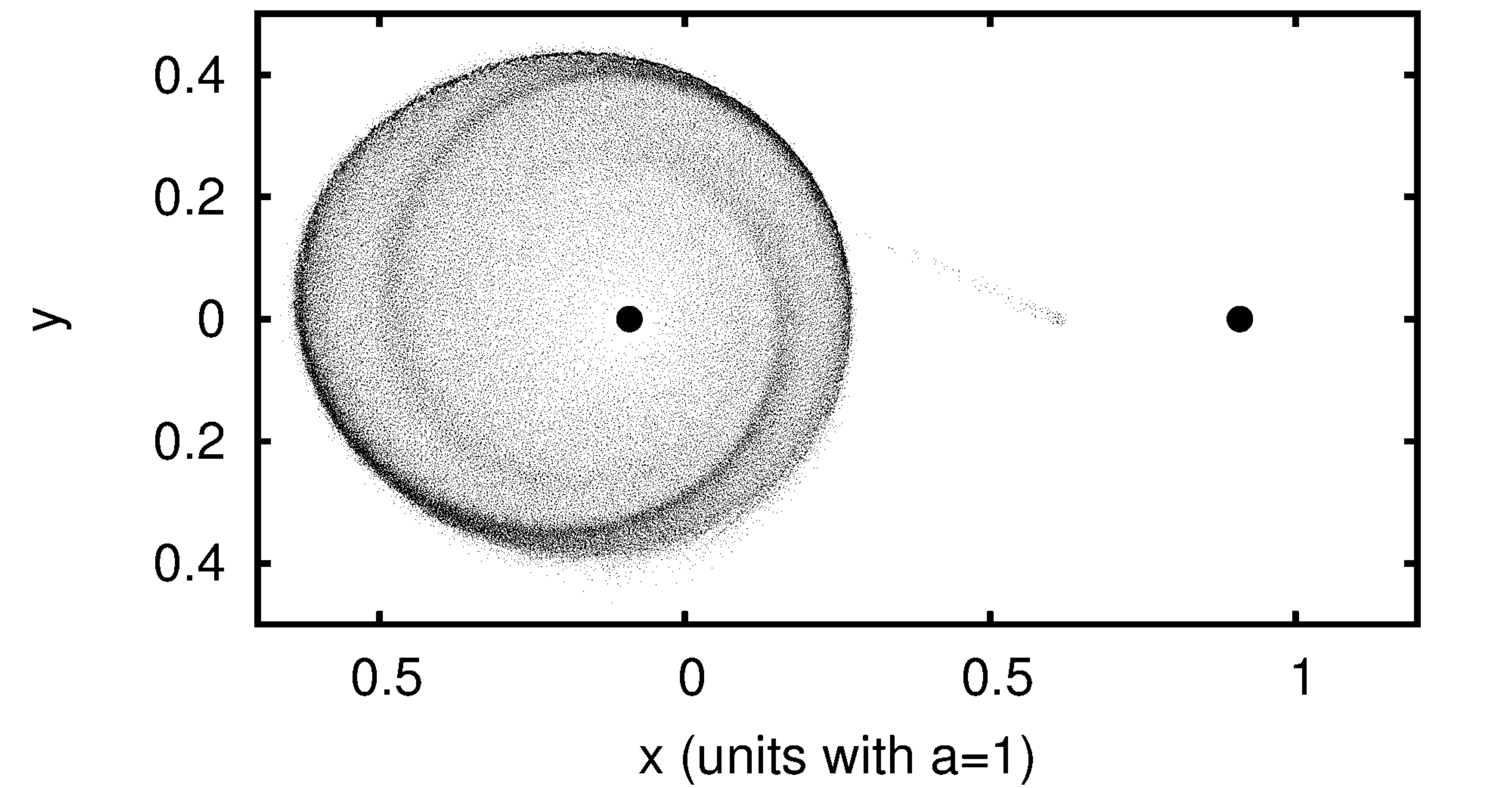}
 \caption{\label{diskplot} Projection of SPH particles onto the
   orbital plane of a double white dwarf system with $q=1/10$, the
   donor and accretor are represented as point masses (the bigger
   dots) and an accretion disk is present around the accretor. The
   coordinates are in units of the separation $a$ between the two
   stars. In this SPH simulation particles were continuously added at
   the inner Lagrange point of the Roche potential. These particles
   have formed the accretion disk around the accretor. This is a
   picture of the 390th orbit after the mass transfer started.}
\end{figure}

Roche lobe filling stars and accretion disks are not completely
symmetric around their centres of mass. Using equations (\ref{h+}) and
(\ref{hx}) we can calculate the contribution this has to the
gravitational wave signal if we know their mass distributions.  For
this study we use a set of SPH simulations of double white dwarfs at
the onset of mass transfer \citep[taken from][]{Dan2011, Dan2012},
hereafter referred to as RocheSPH. We sample the different mass
combinations as shown in Table~ \ref{SPHtable}.  An example of one
star of an SPH simulation is shown in Fig.~\ref{rocheplot} in such a
way that the non-axisymmetric SPH particles are shown more
prominently. In addition we have used an SPH simulation of an
accretion disc in an ultra-compact binary, kindly provided to us by
Prof. Matt Wood (based on \citealt{Wood2009}). The latter is a $M_{\rm
  donor}/M_{\rm accretor} =q=1/10$ system with an accretion disk
around the accretor formed by adding mass at the inner Lagrange point
of the Roche potential with the two stars treated as point masses
(Fig.~\ref{diskplot}), from here onwards referred to as the
DiskSPH. We assume the mass in the disk is $10^{-5}$ of the mass of
the accretor.

All simulation results are snapshots at some point in the evolution of
the binary. Because of this, we can not take all movement of the
system into account, so we assumed that these mass distributions
rotate with fixed angular speeds around their centres of mass. Now
equations \ref{h+}, \ref{hx} and \ref{strainamp} can be used to
calculate the GW strain amplitude. 

\subsection{The results}

\begin{figure}
\includegraphics[height=\columnwidth,angle=-90]{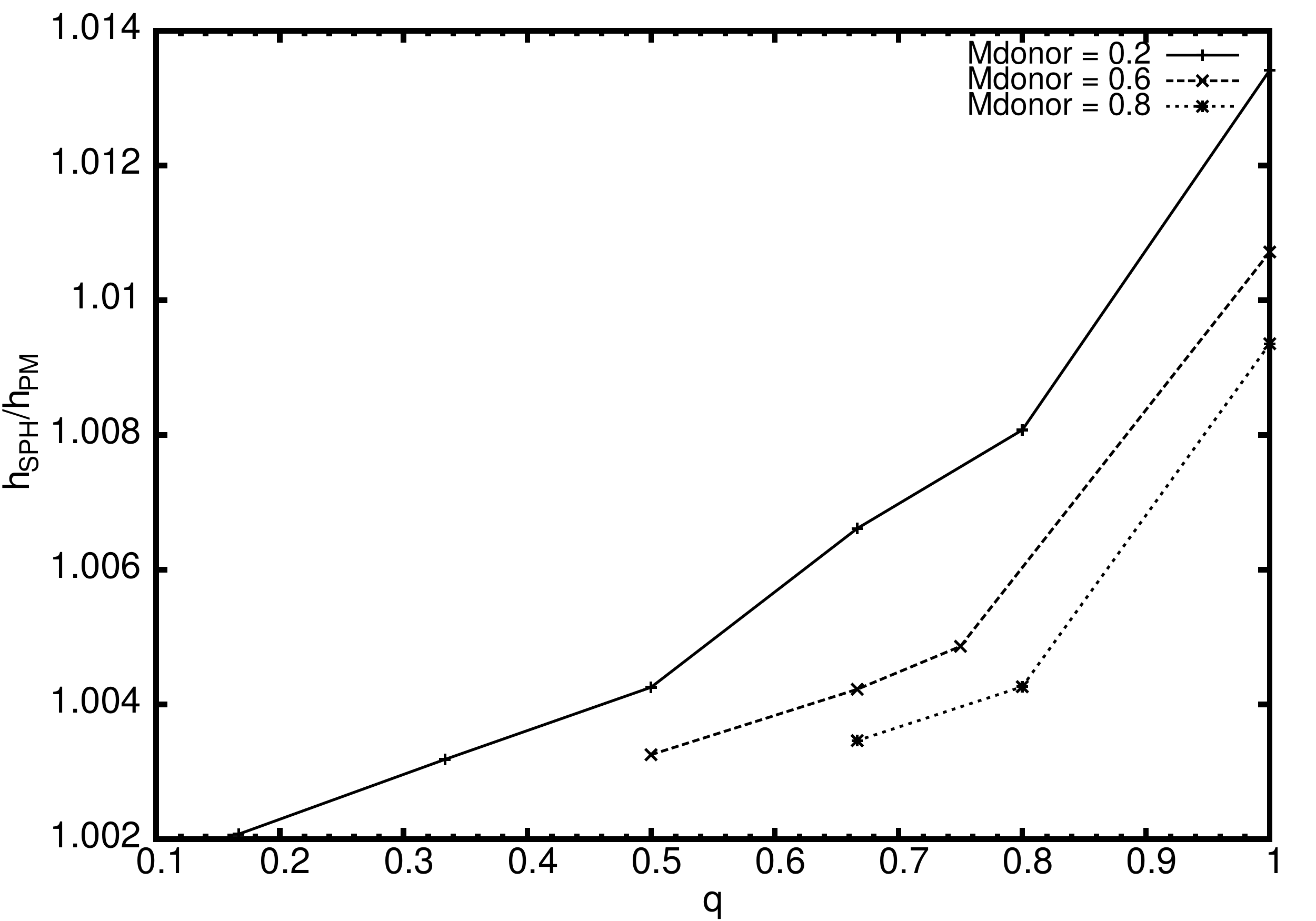}
\caption{\label{rochevary} Ratio of the GW strain amplitude
  (equation \ref{strainamp}) from the RocheSPH systems
  (Fig. \ref{rocheplot}) and corresponding the two point mass
  systems as a function of the mass ratio, for different donor
  masses. See also Table~\ref{SPHtable}}
\end{figure}

The differences of GW strain amplitude for the DiskSPH and all
RocheSPH systems compared to two point systems are listed in Table
\ref{SPHtable}. The DiskSPH results shows a deviation at the $10^{-5}$
level, while the RocheSPH results, depending on the mass of the
deformed donor star and the mass ratio $q$, range between 0.2 and 1.3
per cent. The results of the RocheSPH calculations as function of mass
ratio $q$ and donor mass are shown graphically in
Fig.~\ref{rochevary}. The deviations are largest for equal mass
systems, but more interesting, our results clearly show the fact that
lower donor masses are more deformed than more massive donors at the
same mass ratio. 

\begin{table}
 \caption{\label{SPHtable}Ratio of GW strain amplitude of SPH systems
 and two point mass systems. The mass of the accretion disk in DiskSPH
 was taken to be $10^{-5}M_{\rm accretor}$.}
\begin{center}
 \begin{tabular}{lcccc}
  \hline
  SPH simulation & $M_{\rm donor}$ & $M_{\rm accretor}$ & $q$ & $\frac{h_{\rm SPH}}{h_{\rm PM}}$\\
 &  $(M_\odot)$ & $(M_\odot)$ & &\\ 
  \hline
  DiskSPH &   & & 0.1 & 1.000010\\
  RocheSPH & 0.2 & 0.2  & 1.0  & 1.013418 \\
  RocheSPH & 0.2 & 0.25 & 0.8   &  1.008075 \\
  RocheSPH & 0.2 & 0.3  & 0.667 &  1.006612\\
  RocheSPH & 0.2 & 0.4  & 0.5   &  1.004254\\
  RocheSPH & 0.2 & 0.6  & 0.33  & 1.003183 \\
  RocheSPH & 0.2 & 1.2 & 0.167  & 1.002072 \\
  RocheSPH & 0.6 & 0.6  & 1.0   & 1.010717 \\
  RocheSPH & 0.6 & 0.8  & 0.75  & 1.004861 \\
  RocheSPH & 0.6 & 0.9  & 0.667 & 1.004226 \\
  RocheSPH & 0.6 & 1.2  & 0.5   & 1.003254 \\
  RocheSPH & 0.8 & 0.8  & 1.0   & 1.009355 \\
  RocheSPH & 0.8 & 1.0  & 0.8   & 1.004264 \\
  RocheSPH & 0.8 & 1.2  & 0.667 & 1.003461 \\
  \hline
 \end{tabular}
\end{center}
\end{table}

\begin{figure}
 \includegraphics[keepaspectratio, height= \linewidth,angle=-90]{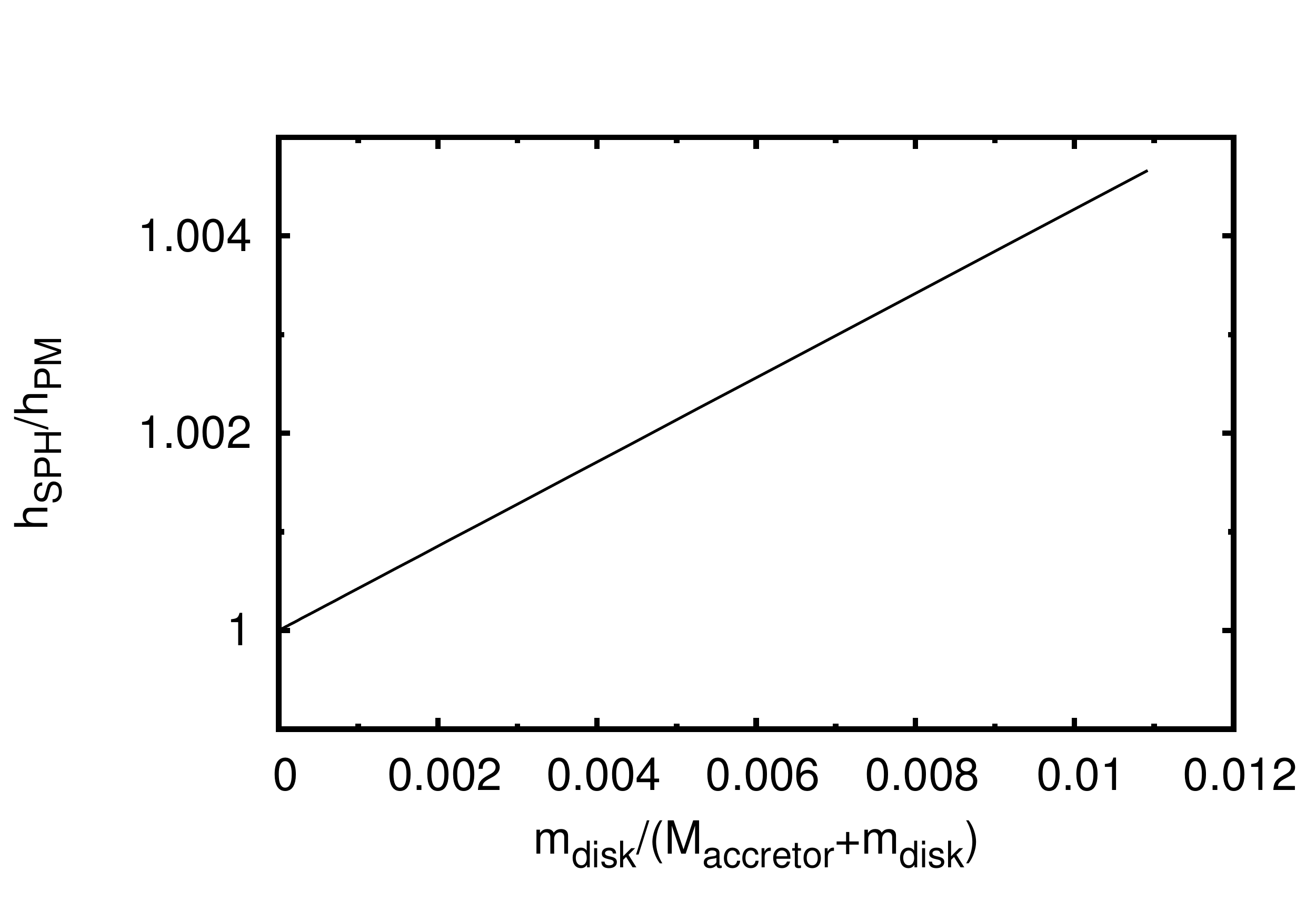}
\caption{\label{diskvary} Ratio of the GW strain amplitude (equation
  \ref{strainamp}) from the DiskSPH system (Fig. \ref{diskplot}) and
  the two point mass system. The $x$-axis shows ratio of disk mass to
  the accretor plus disk mass. The donor is considered to be a point
  mass, so at $m_{\rm disk}=0$ the system is equal to the two point
  system.}
\end{figure}

To explore the influence of an accretion disk in somewhat more detail,
 we varied the mass of the accretion disk compared to the mass in the
 two stars (Fig.~\ref{diskvary}). As can be expected, the deviation
 from the point mass approximation scales linearly with the disk mass,
 which has to be unrealistically high (1 per cent of the total mass)
 in order to get near the effect of the deformed donor star discussed
 above. The deviation for a system with Roche-lobe filling donor plus
 an accretion disk is approximately the sum of the RocheSPH and DiskSPH
 result (see above) and thus for realistic disk masses is dominated by
 the deformation of the donor.

\section{Conclusions}\label{section conclusions}

We have calculated the deviation of the gravitational wave signal of
finite size and non-spherical white dwarfs to that of point masses,
which is usually assumed. For any finite size, non-relativistic
axisymmetric body in circular orbit, the result is exactly the same.
For semi-detached white dwarf binaries, we find that an accretion disk
of reasonable mass changes the gravitational wave signal at the level
of $10^{-4}$ to $10^{-3}$, small but still significantly larger than
errors due to the neglect of post-Newtonian corrections in the
calculation of the signal. Deformations due to filling the Roche lobe
of semi-detached binaries increase for mass ratios closer to unity for
fixed donor mass and are distinctly stronger at fixed mass ratio for
lower mass donors and can in the most extreme cases be of order 1 per
cent. This in principle will change the frequency evolution and the
accuracy with which the parameters can be determined
(e.g. \citealt{blanchet1995}, \citealt{cutler1994}), so calculations
in which accuracies of better than one per cent are needed should take
the finite size into account. Also, the different strength of the
gravitational wave angular momentum losses will affect the mass
transfer rate and stability of the mass transfer in semi-detached
systems (e.g. \citealt{marsh04}). However, the level of deviation we
found shows that the calculations presented in the literature on the
expected signals of (verification) binaries for LISA, are essentially
unaffected: for monochromatic sources, the amplitude of the signal
will be slightly different, but indistinguishable from sources with
slightly higher (chirp) masses and/or smaller distances, properties
that are much more uncertain than one per cent.  For systems with
measurable period derivatives, the degeneracy between (chirp) mass and
distance is broken, but the deviations we found can still only be
detected if there is independent extremely accurate measurement of the
(chirp) mass of the binary. For a LISA-like detector such independent
mass estimates, if available, are typically accurate at the 10 per
cent level at best (e.g. \citealt{littenberg2011}). The detailed
evolution (and possible merger of the system) will be affected as
well, but the calculations of this phase (e.g. \citealt{Dan2012})
already take the finite size of the stars into account. We therefore
conclude that in the majority of cases, the use of the point mass
approximation is well justified.

\section*{Acknowledgments}
We thank Matt Wood for sharing the results of his SPH simulations
with us and the anonymous referee for comments that greatly improved
the paper.

\bsp

\label{lastpage}


\begin{thebibliography}{99}
\bibitem[\protect\citeauthoryear{Amaro-Seoane et al.}{2012}]{eLISA} Amaro-Seoane, P., 
Aoudia, S., Babak, S., et al.\ 2012, arXiv:1202.0839 
\bibitem[\protect\citeauthoryear{Blanchet et al.}{1995}]{blanchet1995} Blanchet, L., Damour, 
T., Iyer, B.~R., Will, C.~M., \& Wiseman, A.~G.\ 1995, Physical Review Letters, 74, 3515
\bibitem[\protect\citeauthoryear{Blaut}{2011}]{blaut2011}  Blaut, A., 2011, PRD, 83, 3006
\bibitem[\protect\citeauthoryear{Brown et al.}{2011}]{brown2011}  Brown, W.R., Kilic, M., Hermes, J. J., Allende Prieto, C., Kenyon,  S.J., Winget, D. E., 2011, ApJ, 737, 23
\bibitem[\protect\citeauthoryear{Cutler \& Flanagan}{1994}]{cutler1994} 
Cutler, C., Flanagan, E. E., PRD, 49, 2658
\bibitem[\protect\citeauthoryear{Dan et al.}{2011}]{Dan2011} 
Dan M., Rosswog S., Guillochon J., Ramirez-Ruiz E., 2011, ApJ, 737, 89 
\bibitem[\protect\citeauthoryear{Dan et al.}{2012}]{Dan2012} Dan M.,
  Rosswog S., Guillochon J., Ramirez-Ruiz E., 2012, arXiv:1201.2406
\bibitem[\protect\citeauthoryear{Eggleton}{1983}]{eggl} Eggleton P.P., 1983, ApJ, 268, 368.
\bibitem[\protect\citeauthoryear{Evans et al.}{1987}]{e1} Evans, C.R., Iben, Jr., I., Smarr, L. 1987 ApJ, 323,129
\bibitem[\protect\citeauthoryear{Lipunov et al.}{1987}]{l1} Lipunov, V.M., Postnov, K.A., Prokhorov, M.E. 1987, A\&A, 220, 135
\bibitem[\protect\citeauthoryear{Hils et al.}{1990}]{h1} Hils, D., Bender, P.L., 2000, ApJ, 537, 334
\bibitem[\protect\citeauthoryear{Littenberg}{2011}]{littenberg2011} Littenberg, T.B., 2011, PRD, 84, 3009
\bibitem[\protect\citeauthoryear{Marsh}{2011}]{marsh2011} Marsh, T.R., 2011, Class. Quantum Grav., 28, 4019
\bibitem[\protect\citeauthoryear{Marsh et~al.}{2004}]{marsh04} Marsh,
  T.R., Nelemans, G., Steeghs, D, 2004, MNRAS, 350, 113
\bibitem[\protect\citeauthoryear{Misner, Thorne, 
\& Wheeler}{1973}]{1973grav.book.....M} Misner, C.~W., Thorne, K.~S., \& Wheeler, J.~A.\ 1973, Gravitation, San Francisco: W.H.~Freeman and Co.
\bibitem[\protect\citeauthoryear{Nelemans et al.}{2001}]{n1} Nelemans, G., Portegies Zwart, S.F., Verbunt, F., Yungelson, L. 2001, A\&A, 368, 939-949
\bibitem[\protect\citeauthoryear{Nelemans et~al.}{2001}]{nyp01}  Nelemans, G., Yungelson, L.~R. and Portegies~Zwart, S.~F., 2001, A\&A,   375, 890--898
\bibitem[\protect\citeauthoryear{Nelemans et~al.}{2004}]{nyp03} --- 2004, MNRAS, 349, 181--192
\bibitem[\protect\citeauthoryear{Price \& Wang}{2008}]{p1} Price, R.H., Wang, Y. 2008 AJP Volume 76, Issue 10, 930-933 
\bibitem[\protect\citeauthoryear{Rindler}{2001}]{rindler} Rindler, W. 2001 Relativity, Oxford University Press
\bibitem[\protect\citeauthoryear{Roelofs et~al.}{2007}]{rgb+07} {Roelofs} G~H~A, {Groot} P~J, {Benedict} G~F, {McArthur} B~E, {Steeghs} D,  {Moracles-Rueda} L, {Marsh} T~R and {Nelemans} G, 2007, ApJ,  666, 1174--1188
\bibitem[\protect\citeauthoryear{Roelofs et~al.}{2010}]{roelofs2010}
  {Roelofs} G~H~A, Rau, A., {Marsh} T~R, {Steeghs} D, {Groot} P~J, and {Nelemans} G,  2010, ApJ,  711, 138
\bibitem[\protect\citeauthoryear{Solheim}{2010}]{solheim2010} Solheim, J-E., 2010,  PASP, 122, 1133
\bibitem[\protect\citeauthoryear{Stroeer \& Vecchio}{2006}]{sv2006}  Stroeer, A., Vecchio, A., 2006, Class. Quantum Grav., 23, 809 
\bibitem[\protect\citeauthoryear{Yoon}{2007}]{y1} Yoon, S.-C., Podsiadlowski, Ph., Rosswog, S., 2007 MNRAS, Volume 380, Issue 3, pp. 933-948.
\bibitem[\protect\citeauthoryear{Wood, Thomas, 
\& Simpson}{2009}]{Wood2009} Wood M.~A., Thomas D.~M., Simpson J.~C., 2009, MNRAS, 398, 2110 
\end{thebibliography}
\end{document}